\documentclass[journal=jacsat,manuscript=article,a4paper,margin=2cm]{achemso}
\usepackage{hyperref} 
\usepackage{algorithm, algorithmicx, algpseudocode}
\usepackage{graphicx}
\usepackage{subcaption} 
\usepackage{tabularx,booktabs,array}
\newcolumntype{Y}{>{\raggedright\arraybackslash}X}

\usepackage{csquotes}
\usepackage{enumitem}
\usepackage{booktabs}   
\usepackage{caption}
\usepackage{amsmath,amssymb}
\usepackage{xurl} 
\usepackage[most]{tcolorbox}

\tcbuselibrary{breakable}

\newcounter{myalg}
\renewcommand{\themyalg}{\arabic{myalg}}

\newenvironment{AlgBox}[2][]{%
  \refstepcounter{myalg}%
  \begin{tcolorbox}[breakable,enhanced,
    colback=white,colframe=black,
    boxrule=0.4pt,arc=0pt,
    left=6pt,right=6pt,top=6pt,bottom=6pt,
    title={Algorithm~\themyalg.\ #2},
    fonttitle=\bfseries,
    #1]%
}{%
  \end{tcolorbox}%
}

\tcbuselibrary{breakable}

\newcommand{\code}[1]{\texttt{\nolinkurl{#1}}}

\usepackage{float}
\usepackage{placeins}   
\usepackage{tabularx}
\usepackage{makecell}
\usepackage{array}
\usepackage{ragged2e}
\usepackage{caption}
\usepackage{adjustbox}

\newcolumntype{L}[1]{>{\RaggedRight\arraybackslash}p{#1}}
\newcolumntype{C}[1]{>{\Centering\arraybackslash}p{#1}}

\newcommand{\casefig}[1]{%
  \includegraphics[height=0.42\textheight,keepaspectratio]{#1}%
}

\usepackage{makecell}
\usepackage{adjustbox}
\usepackage{placeins}

\usepackage{pdflscape}

\newcommand{\CodeUnderscore}{\textunderscore\allowbreak}
\renewcommand{\code}[1]{\begingroup\ttfamily\small\let\_\CodeUnderscore #1\endgroup}

\author{Yu-Chien Huang}
\affiliation{Department of Technology Application and Human Resource Development, National Taiwan Normal University, Taipei 106, Taiwan}

\author{Dennis Chung-Yang Huang}
\email{dcyhuang@okstate.edu}
\affiliation{
Department of Chemistry, Oklahoma State University, 107 Physical Sciences I, Stillwater, Oklahoma 74078, United States
}
\alsoaffiliation{
Institute for Chemical Reaction Design and Discovery (WPI-ICReDD), Hokkaido University, Kita 21, Nishi 10, Kita-ku, Sapporo, Hokkaido 001-0021, Japan
}

\author{Yun-Cheng Tsai}
\email{pecu@ntnu.edu.tw}
\affiliation{Department of Technology Application and Human Resource Development, National Taiwan Normal University, Taipei 106, Taiwan}

\title{Automated Analysis of DFT Output Files for Molecular Descriptor Extraction and Reactivity Modeling}

\abbreviations{DFT, MLR, LFER, NBO}

\begin{document}


\begin{abstract}
Understanding the relationship between molecular structure and chemical reactivity or properties is fundamental to rational molecular design. Linear free energy relationships (LFERs), particularly Hammett analysis, have long served as powerful tools in organic chemistry. Recently, these approaches have been enhanced by the incorporation of computationally derived parameters, enabling broader applicability across diverse molecules and reactions. To facilitate and scale this process, we present \emph{DFTDescriptorPipeline}, a fully automated workflow for extracting quantum chemical descriptors from Gaussian log files and constructing structure--property/reactivity relationships using multivariate linear regression (MLR) models. We validate the workflow across four case studies, including photoswitchable molecules and catalytic reactions. In each case, the models provide interpretable results, demonstrating the versatility of this approach and relevance to a wide range of chemical contexts. We anticipate that this platform will serve as a generalizable framework for integrating quantum chemical calculations into data-driven molecular design.
\end{abstract}

\section{Introduction}
\begin{figure}[htbp]
  \centering
  \includegraphics[width=\linewidth]{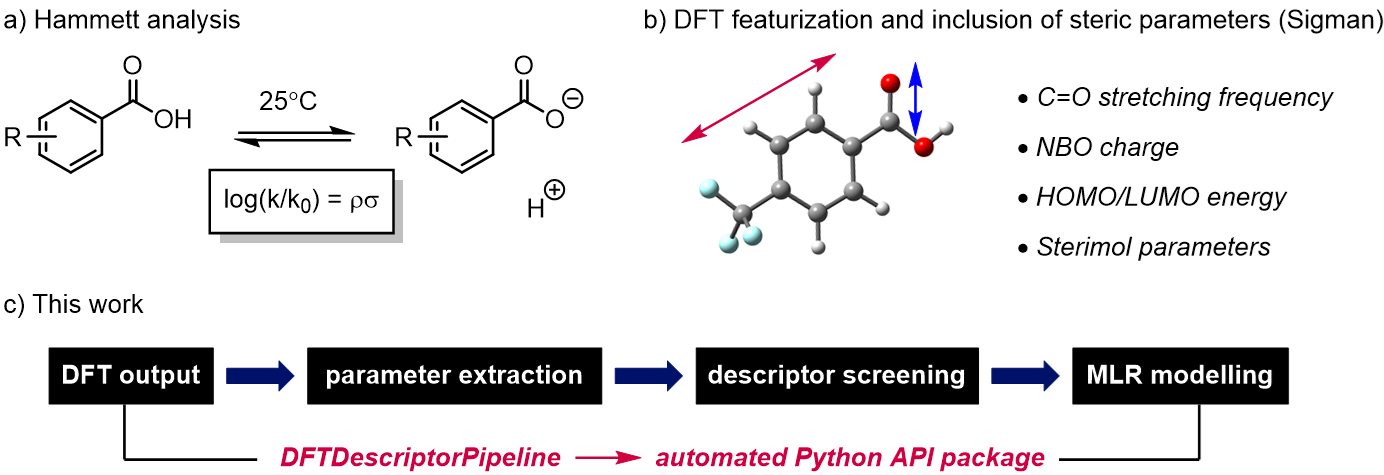}
  \caption{%
  (a) Classical Hammett analysis linking substituent constants to reactivity. 
  (b) DFT-based featurization incorporating steric and electronic parameters. 
  (c) The present work: \emph{DFTDescriptorPipeline} automates descriptor extraction, screening, and multivariate linear regression modeling.}
  \label{fig:workflow}
\end{figure}

Bridging molecular structures with their properties or reactivities has long been a central goal in chemistry. In organic chemistry, linear free energy relationships (LFERs) provide a quantitative framework for understanding how structural variations influence reaction rates or selectivity.\cite{Wells1963}
A classic example is Hammett analysis, in which the acidity of substituted benzoic acids, expressed as the substituent constant ($\sigma$ value), serves as a surrogate parameter for the electronic properties of arenes (Figure~\ref{fig:workflow}a)\cite{Hammett1937}. 
The strength of this approach lies in its ability to infer transition-state energetics (activation barriers) from readily measurable ground-state characteristics. As one of the most influential tools in physical organic chemistry, Hammett analysis has been widely applied to correlate the electron-donating or -withdrawing nature of aromatic substituents with chemical reactivity, selectivity, and other properties.\cite{Jaffe1953}

However, traditional substituent constants are experimentally derived and often unavailable for structurally complex arenes. Moreover, Hammett analysis does not account for steric effects, which can critically influence the catalytic reactivity and selectivity.\cite{Taft1952,Taft1953}
To address these limitations, Sigman and coworkers introduced a novel approach in 2016 that leverages computationally derived parameters (Figure~\ref{fig:workflow}b)\cite{Santiago2016,Santiago2018}
Specifically, they employed density functional theory (DFT) to calculate electronic descriptors (DFT featurization) and incorporated Sterimol values to quantify steric effects.\cite{Verloop1976,Brethome2019}
This method is computationally inexpensive, requiring calculations only for the varying substituents and no transition-state modeling. Using this approach, they constructed multivariate linear regression (MLR) models that successfully correlated reactant structures with the selectivity of catalytic transformations. More recently, one of us adapted this methodology to model the thermal half-lives of \textit{N}-aryl-substituted indigo photoswitches, demonstrating the utility of MLR models in establishing structure–property relationships for functional molecules.\cite{Jaiswal2024}

This modern form of Hammett analysis consists of three key steps: 1) parameter extraction from DFT output files, 2) descriptor tabulation, and 3) construction of MLR models (Figure~\ref{fig:workflow}c). In previous studies, these steps were carried out manually, limiting scalability to larger datasets. Automation of steps 1 and 2 has been reported by Doyle (\emph{Auto-QChem})\cite{Zuranski2022}, Paton (\emph{AQME})\cite{AlegreRequena2023}, and Sigman (\emph{Get\_Properties})\cite{Haas2024} groups. Although these tools have shown great value in data chemistry by connecting computation and molecular parametrization to downstream tasks, a workflow that further integrates LFER model construction remains elusive.\cite{Luchini2024,Lee2024}

We envision a platform that enables users to upload DFT output files along with experimental data, automatically extract relevant molecular descriptors, and identify optimal MLR models. This tool should ideally require minimal coding interventions, thereby lowering the barrier to adoption for general scientists. Herein, we report the development of~\emph{DFTDescriptorPipeline}, an open-source Python package designed for automated parameter extraction and reaction modeling, and demonstrate its utility through four case studies. It is expected that this platform will add to the toolbox for connecting data science and physical organic chemistry.\cite{Crawford2021,Williams2021}

\section{Parametrization}
\label{sec:parametrization}
We implement a fully automated parametrization workflow via 
\url{https://github.com/peculab/DFTDescriptorPipeline},
which extracts quantum-chemical descriptors from Gaussian output (\texttt{.log}) files and aligns them with experimental identifiers.
The process comprises three stages.

\begin{itemize}
  \item \textbf{Descriptor extraction} from Gaussian log files (HOMO/LUMO, dipole, isotropic polarizability, and NBO-based features defined by anchor atoms).
  \item \textbf{Steric quantification} via Sterimol parameters ($L$, $B_1$, $B_5$) computed along the C1--C2 axis.
  \item \textbf{Feature aggregation} into a unified \texttt{pandas} table with substituent-aware prefixes for downstream modeling.
\end{itemize}

\begin{figure}[htbp]
  \centering
  \includegraphics[width=\linewidth]{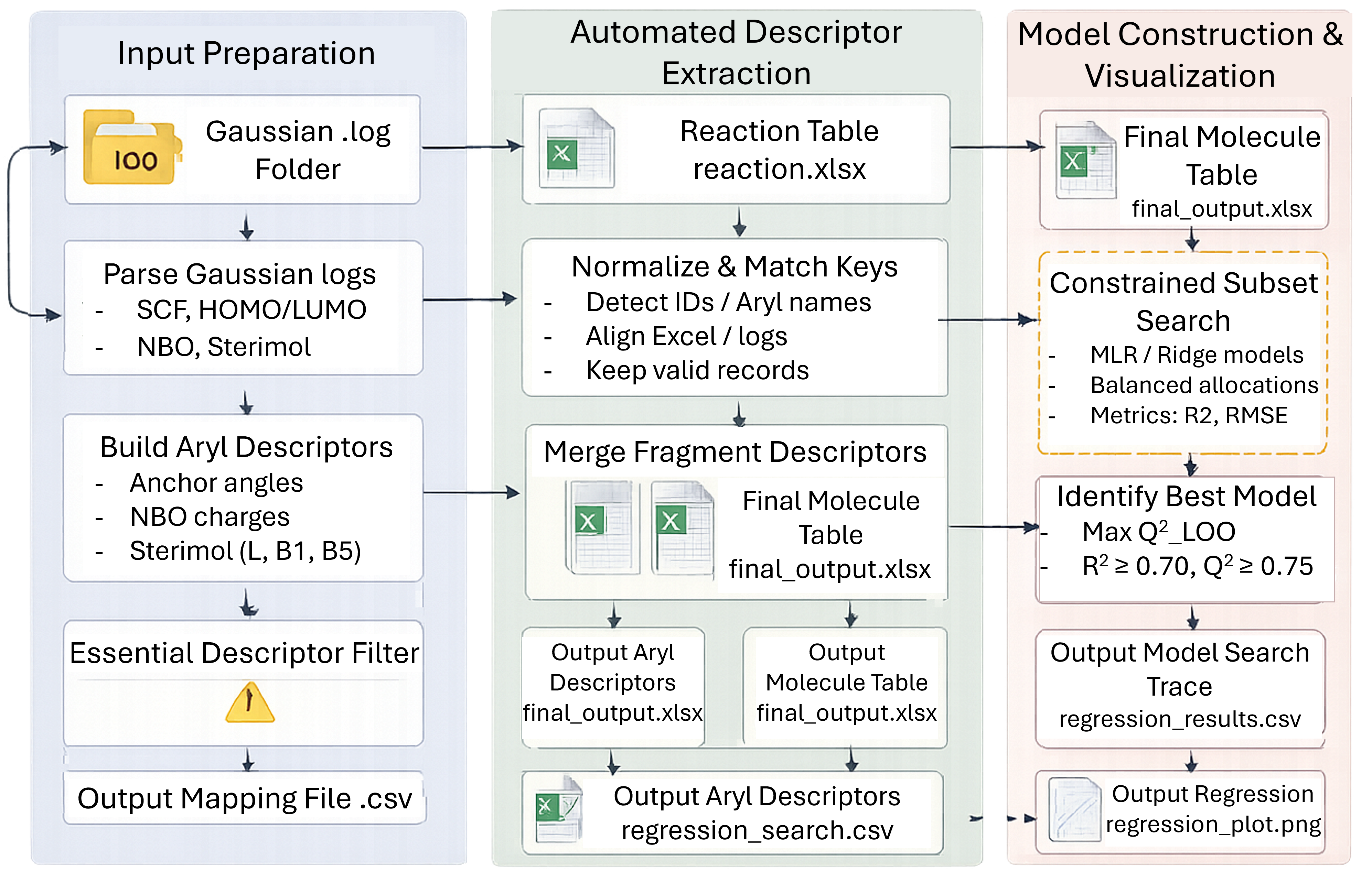}
  \caption{End-to-end pipeline overview of \emph{DFTDescriptorPipeline}: input preparation, automated descriptor extraction,
  substituent matching, model construction, and visualization.}
  \label{fig:flowchart}
\end{figure}

This end-to-end process ensures reproducible and scalable descriptor generation across diverse reaction systems.
A schematic overview of the complete pipeline is provided in Figure~\ref{fig:flowchart}.

Detailed extraction procedures, anchor-based NBO descriptors, Sterimol geometry processing, and the final feature aggregation framework are described in Section~\emph{Descriptor Extraction}.

\paragraph{Anchor indices and Sterimol (implementation details).}
Anchor atoms (a–g) are inferred from the NBO summary by locating the O–H
bond(s) of the acid, the carbonyl carbon (C1), and the adjacent aryl carbon
(C2). Steric descriptors are computed after removing atoms a, b, and d from
the final Standard orientation geometry; C1–C2 defines the Sterimol axis.
Bondi radii are used with a hydrogen adjustment (H = 1.09\,\AA{}). The
resulting $L$, $B_1$, and $B_5$ values are stored as
\texttt{Ar\_Ster\_L}, \texttt{Ar\_Ster\_B1}, \texttt{Ar\_Ster\_B5}.


\section{Descriptor Extraction}
\label{sec:descriptor_extraction}

For each molecule, the pipeline systematically parses Gaussian log files to obtain a comprehensive set of quantum chemical descriptors relevant to structure--reactivity modeling.
The extraction is implemented in a module, which integrates regular-expression parsing, numerical computation, and error handling to ensure robustness across varied log file formats (\href{https://github.com/peculab/DFTDescriptorPipeline}{\texttt{https://github.com/peculab/DFTDescriptorPipeline}}).

\subsection{Frontier Orbital Energies (HOMO/LUMO)}
Frontier orbital energies are extracted from the final converged self-consistent field (SCF) results, specifically the highest occupied (HOMO) and lowest unoccupied (LUMO) molecular orbital energies reported at the end of the electronic-structure calculation. These descriptors reflect the electron-donating and electron-accepting tendencies of substituents.

\subsection{Dipole Moment}
The field-independent dipole magnitude is parsed from the ``Dipole moment'' block in the Gaussian output and reported in Debye units. 
This descriptor suggests overall charge distribution within the molecule.

\subsection{Isotropic Polarizability}
Isotropic polarizability is computed as the arithmetic mean of the tensor components from the ``Exact polarizability'' section,
reflecting the molecular responsiveness to external electric fields.

\subsection{NBO Analysis and Anchor Atoms}
\begin{figure}[htbp]
  \centering
  \includegraphics[width=0.25\linewidth]{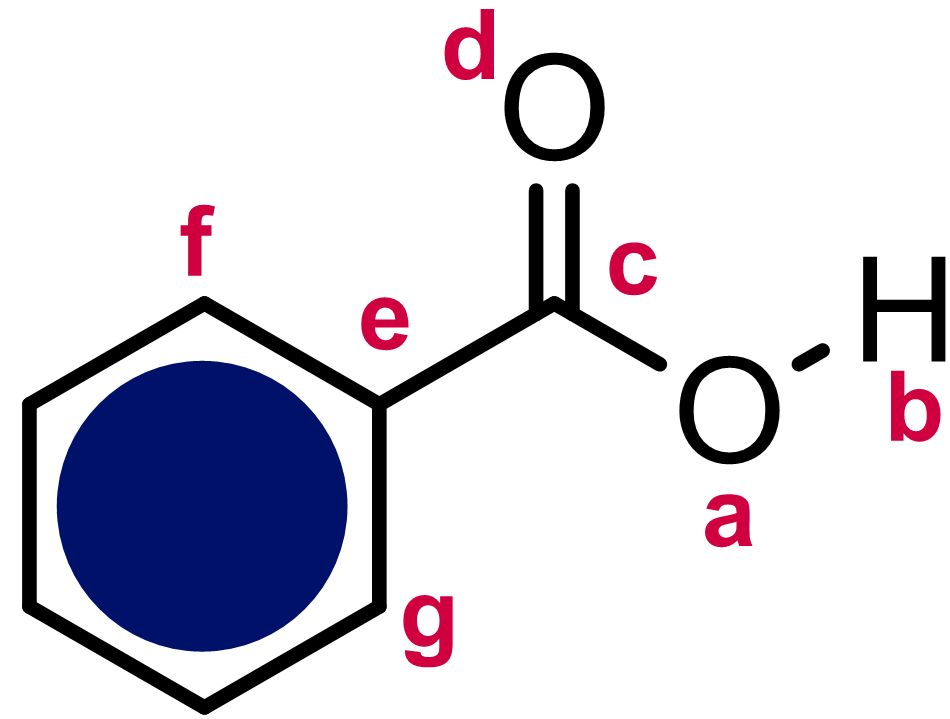}
  \caption{Definition of the seven anchor atoms (a–g) used for NBO analysis.
  Atom c (C1) is the carbonyl carbon; atom e (C2) is the connecting aryl carbon.}
  \label{fig:anchor}
\end{figure}

This module locates the ``Natural Bond Orbitals (Summary)'' block and identifies the O--H bonds of the carboxylic acid moiety, 
which serve as reference points for defining seven characteristic atoms (a--g) (Figure~\ref{fig:anchor}).
Atom c (C1) corresponds to the carbonyl carbon of the carboxylic acid, and atom e (C2) to the connecting carbon on the aromatic ring,
while a, b, d, f, and g represent adjacent atoms and aryl extensions.
These anchor atoms provide consistent geometric references for extracting:
\begin{itemize}
  \item Occupancies and orbital energies of the C1--O and C1--C2 bonds,
  \item Atomic NBO charges on the C1, C2, and O atoms within the carboxyl group,
  \item IR vibrational frequency and intensity of the C=O stretch (in the range of 1800--1900~cm$^{-1}$),
  \item C1--C2 bond length, computed from the final ``Standard orientation'' geometry.
\end{itemize}
All parsing routines are tolerant to format variations and automatically report missing atomic indices.

\subsection{Sterimol Parameter Computation}
\begin{figure}[htbp]
  \centering
  \includegraphics[width=0.25\linewidth]{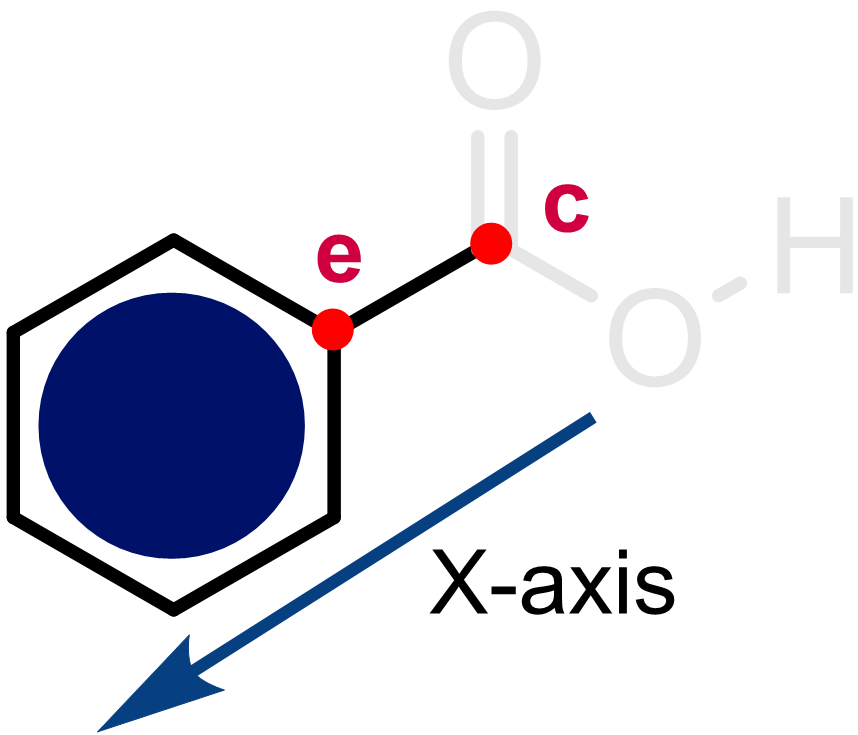}
  \caption{Calculation of Sterimol parameters ($L$, $B_1$, and $B_5$) along the C1--C2 axis. 
  Bondi radii are applied to define atomic boundaries.}
  \label{fig:sterimol}
\end{figure}

To capture steric effects, Sterimol parameters ($L$, $B_1$, $B_5$) are automatically computed using the \texttt{morfeus} package.\cite{Brethome2019}
Prior to computation, atoms a, b, and d are excluded from the geometry, and the C1--C2 bond is designated as the reference X-axis (Figure~\ref{fig:sterimol}).
The structured procedure is as follows:
\begin{enumerate}
  \item \textbf{Anchor Atom Selection:} Indices defining the substituent framework (a--g) are identified from NBO analysis. 
  Atoms c (C1) and e (C2) define the substituent attachment axis.
  \item \textbf{Geometry Filtering:} Atoms a, b, and d are removed from the final ``Standard orientation'' geometry,
  and the filtered coordinates are written to an intermediate \texttt{.xyz} file.
  \item \textbf{Axis Definition and Descriptor Computation:} 
  Using the C1--C2 bond as the principal X-axis, the Sterimol parameters are computed as: 
  $L$ (maximum substituent length), $B_1$ (minimum width), and $B_5$ (maximum width). 
  Atomic radii are assigned using Bondi radii, with adjusted values for hydrogen.
  \item \textbf{Error Handling:} Molecules lacking valid anchor atoms or yielding invalid geometries are skipped. 
  Missing values are set to \texttt{None}, and all exceptions are logged for transparency.
  \item \textbf{Integration into Feature Table:} The resulting Sterimol parameters are appended to the descriptor dataframe 
  with substituent-specific prefixes (e.g., \texttt{Ar1\_Ster\_L}, \texttt{Ar2\_Ster\_B5}).
\end{enumerate}

\subsection{Feature Table Aggregation}
All extracted descriptors (electronic, vibrational, geometric, and steric) are consolidated into a unified tabular representation that serves as the direct input to downstream regression analyses. The aggregation procedure is designed to preserve provenance (i.e., which substituent slot a descriptor originates from), and maintain robustness against partial extraction failures. Concretely, the workflow proceeds as follows:
\begin{enumerate}
  \item \textbf{Feature-table assembly:} For each unique substituent identifier (e.g., Ar, Ar1, Ar2), the extracted descriptors are compiled into a single dataframe, one row per substituent. Column names are systematically prefixed by the corresponding substituent slot to retain traceability after merging at the molecule level.
  \item \textbf{Completeness screening:} A code-defined set of essential descriptors (including NBO charges, HOMO/LUMO energies, IR vibrational frequency/intensity of C=O stretch, and Sterimol parameters) is checked for missing values. Entries lacking any critical fields are excluded to prevent ill-posed regression fits and to improve the stability of cross-validated performance.
  \item \textbf{Non-blocking error reporting:} Extraction anomalies (e.g., missing anchor atoms, failed parsing blocks, or geometry-related exceptions) are recorded in a structured diagnostic report. This design enables post hoc inspection and debugging without interrupting the end-to-end pipeline execution.
  \item \textbf{Interoperable export:} The finalized feature table is exported in common tabular formats suitable for reproducible statistical analysis and for integration with external modeling toolchains.
\end{enumerate}

\subsection{Algorithmic Overview}
\label{sec:algo_overview}

We implement a fully automated pipeline that maps Gaussian log files and a reaction spreadsheet into
a multivariate regression model via descriptor extraction, prefix-based merging, and an exhaustive,
constraint-aware subset search. Formally, let $\mathcal{A}$ be the set of unique aryl identifiers
(keys) appearing in the reaction table, and let $\mathcal{D}=\{\ell_a\}_{a\in\mathcal{A}}$ denote the
corresponding Gaussian \texttt{.log} files. The pipeline constructs an aryl-level descriptor map
$\phi: a \mapsto \mathbf{f}_a \in \mathbb{R}^{p}$, merges descriptors into a molecule-level design
matrix $\mathbf{X}\in\mathbb{R}^{n\times P}$ aligned with the response vector $\mathbf{y}\in\mathbb{R}^{n}$,
and selects the best linear model by maximizing the leave-one-out cross-validated coefficient of determination
$Q^2_{\mathrm{LOO}}$ under group-balance constraints.

\vspace{0.5em}
\begin{AlgBox}{Code-consistent automated extraction and subset model search}
\label{alg:extract_methodology}

\begin{algorithmic}[1]
\Require Reaction table $\mathcal{X}$ (spreadsheet) containing aryl slots $\{\texttt{Ar1},\texttt{Ar2},\ldots\}$ and target $y$;
Gaussian logs $\mathcal{D}=\{\ell_a\}$ indexed by aryl key $a$;
join mode $J\in\{\textsf{single},\textsf{pair},\ldots\}$;
maximum subset size $k_{\max}$;
per-group bounds $[m_g,M_g]$ (default $m_g=1$, $M_g=3$) and a performance filter (e.g., $R^2\ge 0.70$).
\Ensure Merged table $\mathcal{T}$; search results $\mathcal{R}$; best subset $S^\star$ with fitted coefficients $\hat{\boldsymbol{\beta}}$.

\Statex \vspace{0.4em}\textbf{I. Key normalization and aryl-level descriptor map}
\State Extract raw keys from $\mathcal{X}$; normalize tokens (case/spacing tolerant; numeric canonicalization, e.g., \texttt{"101.0"}$\to$\texttt{"101"}).
\State Define the valid aryl set $\mathcal{A}\gets\{a:\exists\,\ell_a\in\mathcal{D}\}$ and remove reaction rows referencing missing keys.
\ForAll{$a\in\mathcal{A}$}
  \State Parse $\ell_a$ to obtain electronic and structural descriptors.
  \Statex \hspace{1.2em} HOMO/LUMO, dipole, polarizability,
  \Statex \hspace{1.2em} NBO anchors/charges, key bond metrics (C1--O, C1--C2), and C{=}O IR (1800--1900\,cm$^{-1}$).
  \State Compute Sterimol descriptors $(L,B_1,B_5)$ along axis C1--C2 with Bondi radii (H adjusted) after removing predefined anchor atoms.
  \State Assemble $\mathbf{f}_a$ and store in an aryl feature table $\mathcal{F}_{Ar}$ (one row per $a$).
\EndFor
\State Essential-feature screening: remove aryl rows with missing values in a predefined core set $\mathcal{E}$ (code-defined).

\Statex \vspace{0.4em}\textbf{II. Prefix-based merge into molecule-level data}
\State Merge $\mathcal{F}_{Ar}$ into $\mathcal{X}$ according to $J$ by prefixing descriptors per aryl slot.
\Statex \hspace{1.2em} e.g., slot \texttt{Ar1} $\Rightarrow$ columns \texttt{Ar1\_*}, slot \texttt{Ar2} $\Rightarrow$ \texttt{Ar2\_*}.
\State Obtain merged table $\mathcal{T}$ and construct modeling data after dropping rows with NA in $\{y\}\cup$ selected predictors.
\Statex \hspace{1.2em} $\mathbf{y}\in\mathbb{R}^{n}$ (target), $\mathbf{X}\in\mathbb{R}^{n\times P}$ (candidate predictors).

\Statex \vspace{0.4em}\textbf{III. Constraint-aware subset search with closed-form leave-one-out (LOO)}
\State Partition predictor indices $\{1,\ldots,P\}$ into groups $\{\mathcal{G}_g\}_{g=1}^{G}$ by prefix family.
\State Define the feasibility constraint for a subset $S\subseteq\{1,\ldots,P\}$:
\Statex \hspace{1.2em}
$\mathcal{C}(S)=\big(|S|=k\big)\ \wedge\ \big(\forall g,\ m_g \le |S\cap \mathcal{G}_g|\le M_g\big).$
\State Initialize record set $\mathcal{R}\gets\emptyset$.
\For{$k=1$ to $k_{\max}$}
  \ForAll{subsets $S\subseteq\{1,\ldots,P\}$ with $|S|=k$}
    \If{$\mathcal{C}(S)$ is false} \State \textbf{continue} \EndIf
    \State Form $\mathbf{X}_S\in\mathbb{R}^{n\times k}$ and fit OLS:
    \Statex \hspace{1.2em}
    $\hat{\boldsymbol{\beta}}_S=(\mathbf{X}_S^\top\mathbf{X}_S)^{-1}\mathbf{X}_S^\top\mathbf{y}$,
    \quad $\hat{\mathbf{y}}=\mathbf{X}_S\hat{\boldsymbol{\beta}}_S$.
    \State Compute $\mathbf{H}=\mathbf{X}_S(\mathbf{X}_S^\top\mathbf{X}_S)^{-1}\mathbf{X}_S^\top$ and $h_{ii}=\mathrm{diag}(\mathbf{H})$.
    \State Closed-form LOO prediction for each sample $i$:
    \State \State $\hat{y}^{(-i)}_i = y_i - \dfrac{y_i-\hat{y}_i}{1-h_{ii}}.$\\
    \State Compute $R^2$, $Q^2_{\mathrm{LOO}}$, and $\mathrm{RMSE}_{\mathrm{LOO}}$: 
    \State \State $R^2=1-\dfrac{\sum_i (y_i-\hat{y}_i)^2}{\sum_i (y_i-\bar{y})^2}$, 
    \quad
    \State \State $Q^2_{\mathrm{LOO}}=1-\dfrac{\sum_i (y_i-\hat{y}^{(-i)}_i)^2}{\sum_i (y_i-\bar{y})^2}$, 
    \quad
    \State \State $\mathrm{RMSE}_{\mathrm{LOO}}=\sqrt{\dfrac{1}{n}\sum_i (y_i-\hat{y}^{(-i)}_i)^2}$. \\
    \quad
    \If{$R^2 < 0.70$} \State \textbf{continue} \EndIf
    \State Append $(S,\hat{\boldsymbol{\beta}}_S,R^2,Q^2_{\mathrm{LOO}},\mathrm{RMSE}_{\mathrm{LOO}})$ to $\mathcal{R}$.
  \EndFor
\EndFor
\State Save $\mathcal{R}$ to \texttt{regression\_search\_results.csv}.
\State Select the best subset with tie-breaking:
\Statex \hspace{1.2em}
$S^\star = \arg\max_{(S,\cdot)\in\mathcal{R}} Q^2_{\mathrm{LOO}}$ (ties broken by minimal $\mathrm{RMSE}_{\mathrm{LOO}}$).
\State Generate the final parity plot and report the fitted regression equation for $S^\star$.
\end{algorithmic}

\end{AlgBox}

\section{Modeling}

The modeling module of the \emph{DFTDescriptorPipeline} automatically correlates the cleaned
descriptor table with experimental reactivity metrics (e.g., $\ln k_{\mathrm{obs}}$ or
$\Delta\Delta G^{\ddagger}$) using multivariate linear regression (MLR). All steps are executed
in open-source Python code, ensuring full reproducibility from descriptor extraction to
publication-ready plots.

\paragraph{Data cleaning and effective dataset size.}
Before model fitting, rows with missing entries in any descriptor or target column are removed
programmatically. The resulting effective dataset
size ($n_\mathrm{eff}$) can therefore be smaller than the raw spreadsheet count ($n_\mathrm{raw}$),
as only molecules with complete descriptor vectors are used for regression. All reported
$R^2$, $Q^2_\mathrm{LOO}$, and RMSE values correspond to this filtered dataset.

\paragraph{Descriptor grouping.}
Each descriptor column carries an explicit aryl prefix (\texttt{Ar1\_}, \texttt{Ar2\_}, …),
allowing substituent-specific grouping. Features are collected into groups
based on these prefixes to ensure that each substituent contributes at least one descriptor.
The algorithm constructs balanced models by enforcing per-group bounds
(default 1–3 features per group) while varying the total descriptor count up to the user-defined
maximum.

\paragraph{Regression search logic.}
For each feasible feature allocation, all possible combinations are enumerated and a linear model is
fitted to the target variable. Candidate models are screened and ranked through:
\begin{enumerate}[label=(\roman*),leftmargin=*,itemsep=0pt,parsep=0pt]
  \item \textbf{In-sample fit.} Models with $R^2<0.70$ are discarded.
  \[
    R^2 = 1 - \frac{\sum_i (y_i - \hat{y}_i)^2}{\sum_i (y_i - \bar{y})^2}
  \]
  \item \textbf{Cross-validation.} Leave-one-out cross-validation (LOOCV):
  \[
    Q^2_\mathrm{LOO} = 1 - \frac{\sum_i (y_i - \hat{y}_{i,-i})^2}{\sum_i (y_i - \bar{y})^2}.
  \]
  \item \textbf{Error metric.} Root-mean-square error (RMSE):
  \[
    \mathrm{RMSE} = \sqrt{\frac{1}{n}\sum_i (y_i - \hat{y}_i)^2}.
  \]
\end{enumerate}

\paragraph{Model selection.}
Models are ranked primarily by $Q^2_\mathrm{LOO}$ and secondarily by RMSE.
Among models of comparable performance, the algorithm favors those with fewer descriptors
and balanced group representation. The final model $M_{\mathrm{best}}$ takes the form
\[
  y = \beta_0 + \sum_{j=1}^{k}\beta_j x_j,
\]
where $x_j$ are standardized descriptors and $\beta_j$ the fitted coefficients.

\paragraph{Visualization.}
Each final model is visualized through a parity plot (predicted vs.\ experimental) annotated with
$R^2$, $Q^2_\mathrm{LOO}$, RMSE, and $n_\mathrm{eff}$, providing an at-a-glance validation of
model reliability.

\section{Case Studies}
\label{sec:case_studies}

To demonstrate a consistent regression-style reporting across datasets with different coupling schemes (single- vs.\ pair-join) and different targets, we applied the proposed workflow to four case studies: (i) thermal back-reaction of azoarene photoswitches, (ii) redox-relay Heck coupling with boronic acids, (iii) thermal back-isomerization of \emph{N}-aryl-\emph{N}$'$-alkylindigo photoswitches, (iv) thermal back-isomerization of  \emph{N},\emph{N}$'$-diarylindigo photoswitches.

For each case, we report the best models found under a constrained subset search with LOOCV-based $Q^2$. We note whether a model meets the predefined pass criteria; if not, we still report the best-performing model(s) for transparency and comparison. Table~\ref{tab:summary} summarizes the best-performing model in each case study, selected by LOOCV $Q^2$ while prioritizing interpretability when performance is comparable. Please refer to the Associated Content in the project repository (\url{https://github.com/peculab/DFTDescriptorPipeline}) for lists of the top-five performing models.

\begin{figure}[htbp]
  \centering
  \includegraphics[width=1\linewidth]{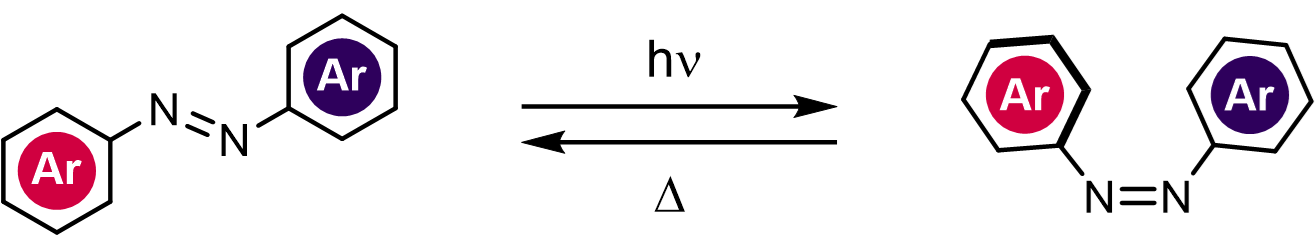}
  \caption{Thermal Back-reaction of Azoarene photoswitches.}
  \label{fig:scheme1}
\end{figure}

\begin{figure}[htbp]
  \centering
  \includegraphics[width=1\linewidth]{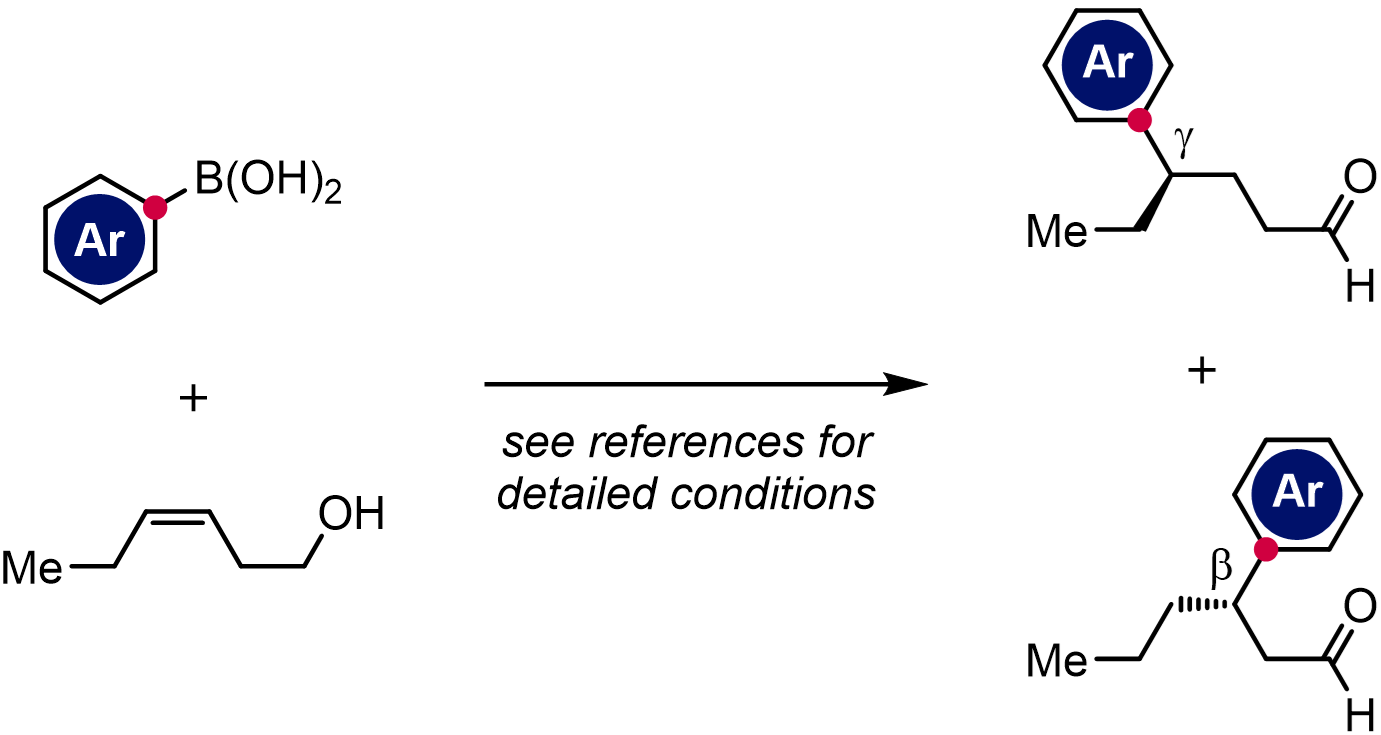}
  \caption{redox-relay Heck coupling with boronic acids.}
  \label{fig:scheme2}
\end{figure}

\begin{figure}[htbp]
  \centering
  \includegraphics[width=1\linewidth]{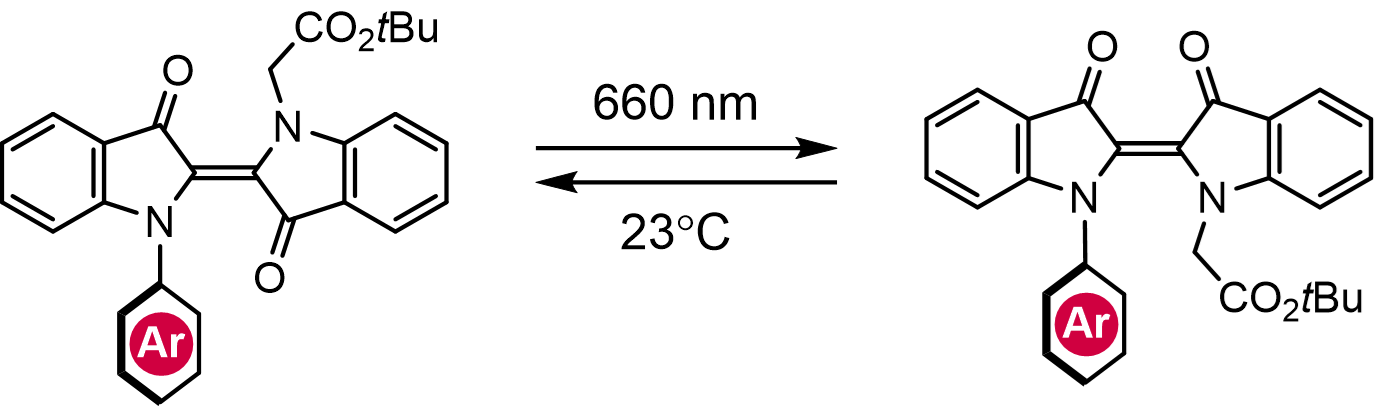}
  \caption{Thermal Back-isomerization of \emph{N}-aryl-\emph{N}$'$-alkylindigo photoswitches.}
  \label{fig:scheme3}
\end{figure}

\begin{figure}[htbp]
  \centering
  \includegraphics[width=1\linewidth]{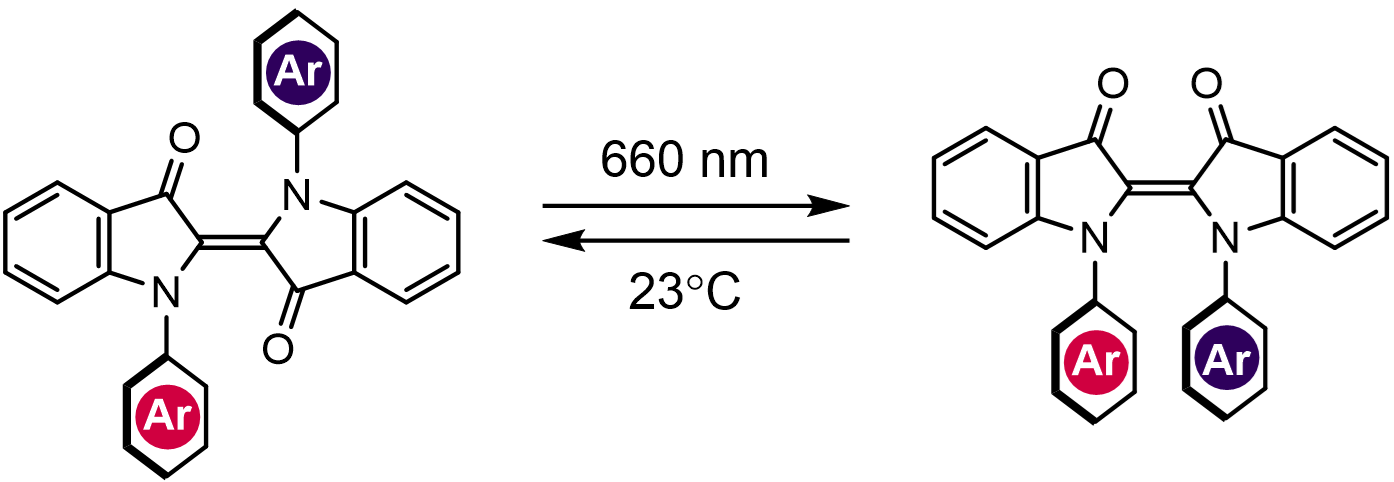}
  \caption{Thermal Back-isomerization of \emph{N},\emph{N}$'$-diarylindigo photoswitches.}
  \label{fig:scheme4}
\end{figure}

\subsection{Thermal Back-Reaction of Azoarene Photoswitches}

Azoarenes are widely used photoswitches whose thermal back-reaction rates are governed by substituent electronics and sterics (Figure~\ref{fig:scheme1}). This azoarene dataset uses a \textbf{pair-join} scheme (\texttt{pair:Ar1+Ar2}), where each reaction entry is formed by combining two aromatic fragments.
We evaluated models with $k=3,4,5$ descriptors. The best LOOCV generalization within this range was obtained at $k=4$ ($R^2=0.647$, $Q^2=0.495$, RMSE $=2.63$, $n=26$), which \textit{does not} meet the pass threshold.

For interpretability, we express the best $k=4$ model as:
\begin{equation}
\ln(k_{\mathrm{obs}}) = 54.05\,x_1 - 0.01\,x_2 - 35.17\,x_3 + 1.32\,x_4 - 18.26,
\end{equation}
where $x_1$ denotes the HOMO energy of Ar$_2$ (\texttt{Ar2\_homo\_hartree}),
$x_2$ denotes the SCF total energy of Ar$_2$ (\texttt{Ar2\_scf\_energy\_hartree}), i.e., the final converged electronic energy reported by the SCF procedure (an output descriptor rather than a user-specified input parameter),
$x_3$ denotes the atomic charge at site $a$ on Ar$_1$ (\texttt{Ar1\_q\_a}),
and $x_4$ denotes the atomic charge at site $g$ on Ar$_2$ (\texttt{Ar2\_q\_g}).

\begin{center}
\casefig{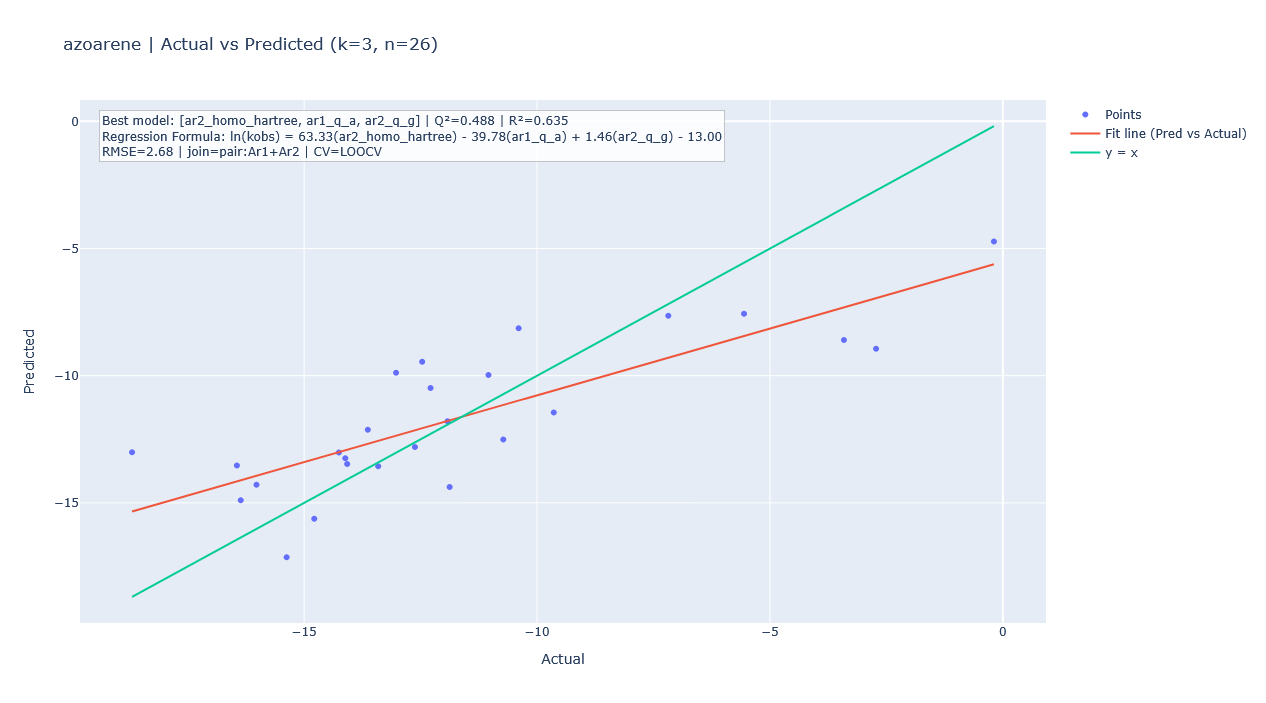}

\vspace{0.1em}
\casefig{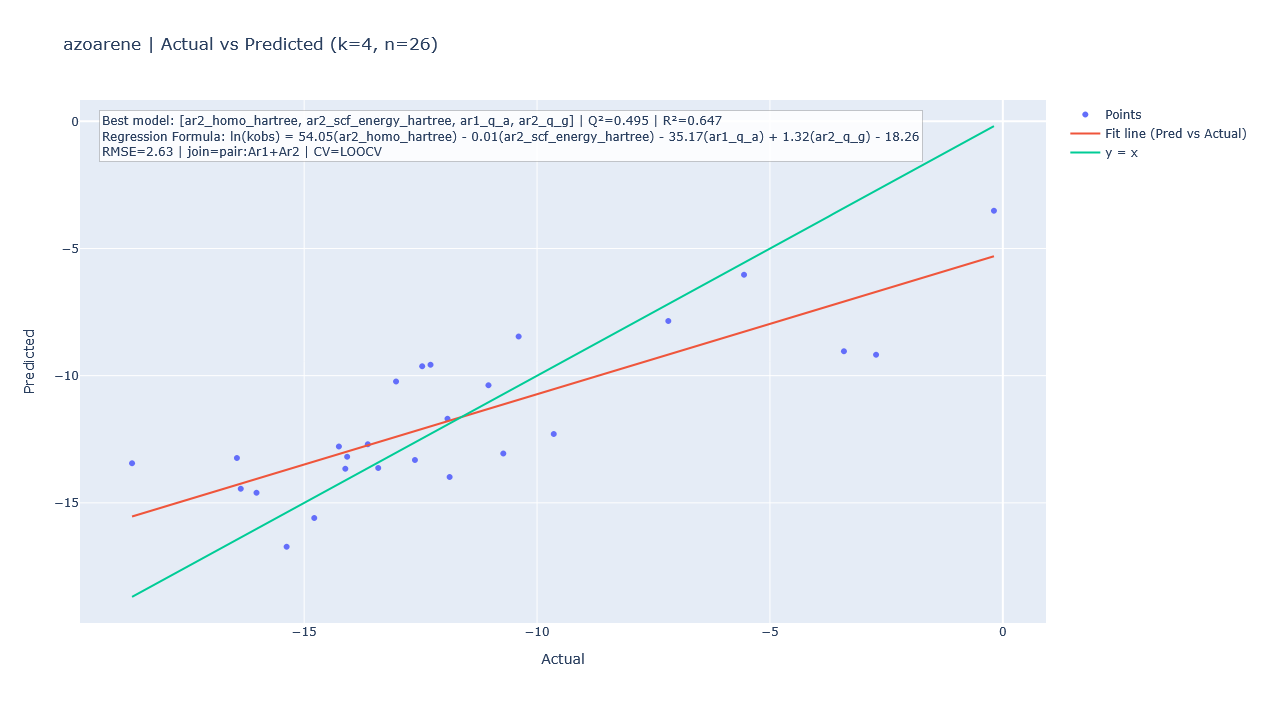}

\vspace{0.1em}
\casefig{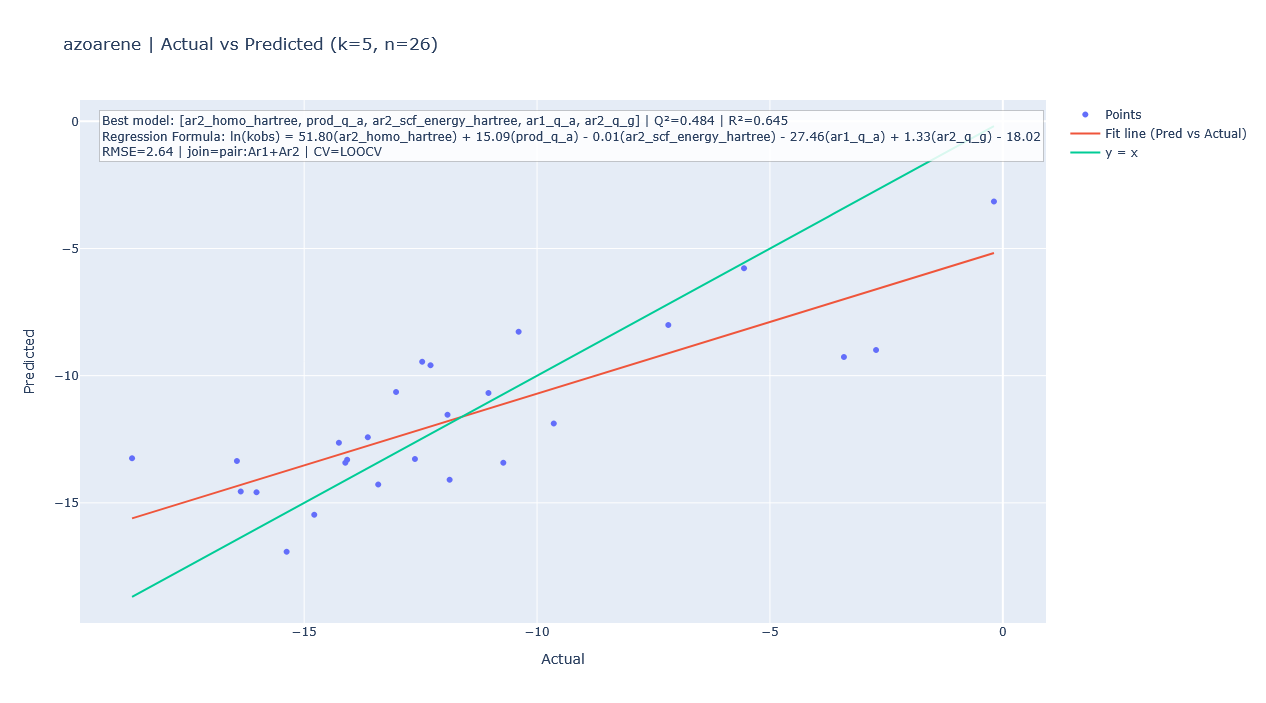}

\captionof{figure}{Azoarene regression plots under \texttt{pair:Ar1+Ar2} join.
(a) $k=3$: $n=26$, $R^2=0.635$, $Q^2=0.488$, RMSE $=2.68$.
(b) $k=4$: $n=26$, $R^2=0.647$, $Q^2=0.495$, RMSE $=2.63$.
(c) $k=5$: $n=26$, $R^2=0.645$, $Q^2=0.484$, RMSE $=2.64$.}
\label{fig:azoarene}
\end{center}

\subsection{Redox-Relay Heck Coupling with Boronic Acids}

To assess transferability, we examined redox-relay Heck couplings between boronic acids and allylic alcohols (Figure~\ref{fig:scheme2}). This Heck Boronic Acids dataset uses a \textbf{single-join} scheme (\texttt{single:Ar}), where each entry maps to one aromatic fragment.
Across $k=3,4,5$, the best LOOCV performance was obtained at $k=5$ ($R^2=0.889$, $Q^2=0.745$, RMSE $=0.223$, $n=17$), which is \textit{near} but still below the pass threshold on $Q^2$.

The best $k=5$ model is:
\begin{equation}
\Delta\Delta G = -2.22\,x_1 + 22.00\,x_2 - 5.97\,x_3 + 45.89\,x_4 + 2.81\,x_5 - 0.34,
\end{equation}
where $x_1,\dots,x_5$ denote atomic charges at five predefined sites on the aromatic fragment (\texttt{q\_*} features),
obtained by mapping each site to an atom index from the final optimized geometry and then extracting the corresponding values from the atomic-charge section of the Gaussian output.

\begin{center}
\casefig{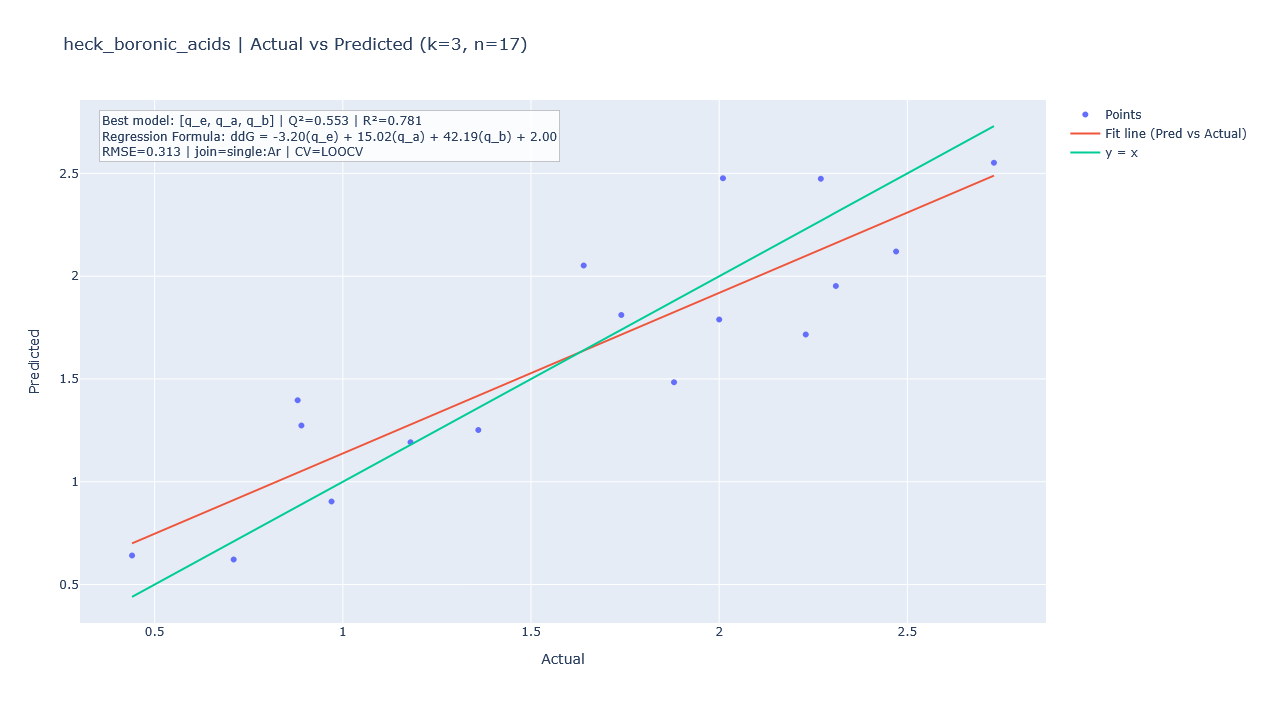}

\vspace{0.1em}
\casefig{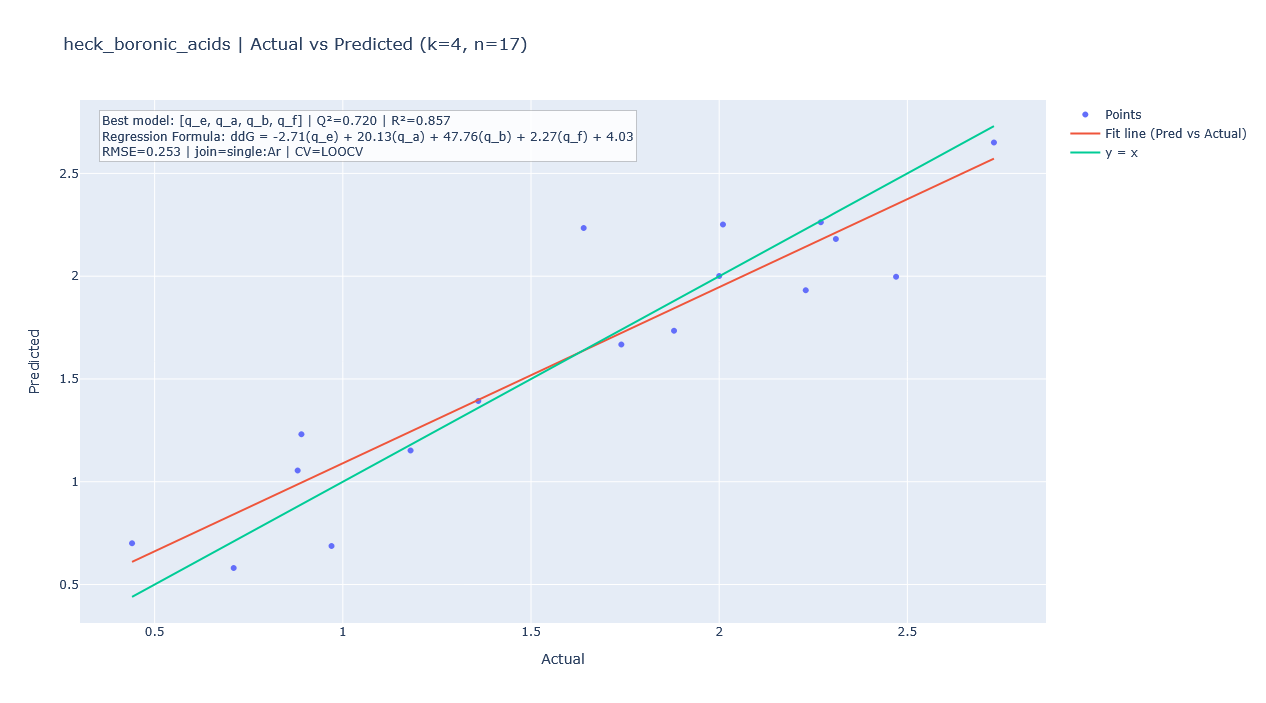}

\vspace{0.1em}
\casefig{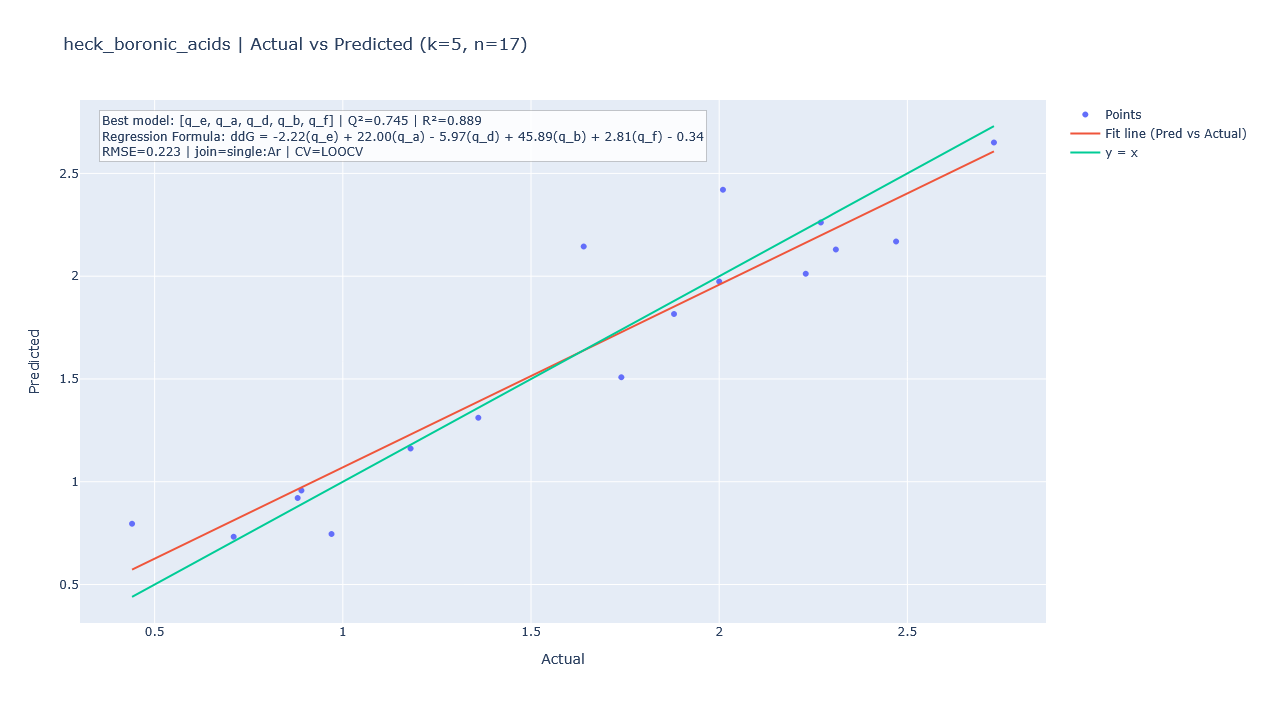}

\captionof{figure}{Heck Boronic Acids regression plots under \texttt{single:Ar} join.
(a) $k=3$: $n=17$, $R^2=0.781$, $Q^2=0.553$, RMSE $=0.313$.
(b) $k=4$: $n=17$, $R^2=0.857$, $Q^2=0.720$, RMSE $=0.253$.
(c) $k=5$: $n=17$, $R^2=0.889$, $Q^2=0.745$, RMSE $=0.223$.}
\label{fig:heck}
\end{center}

\subsection{Thermal Back-isomerization of \emph{N}-aryl-\emph{N}$'$-alkylindigo Photoswitches}

Indigos are an emerging class of red-light photoswitches, and their thermal back-isomerization rates are influenced by  \emph{N}-substituents~\cite{Huang2023}.
We applied the workflow to \emph{N}-aryl-\emph{N}$'$-alkylindigo photoswitches (Figure~\ref{fig:scheme3}).
This Indigo Aryl--Alkyl dataset uses a \textbf{single-join} scheme (\texttt{single:Ar1}) and targets the rate of thermal back-reaction in acetonitrile, denoted as $\ln(k_{\mathrm{obs}})_{\mathrm{MeCN}}$.
Among $k=3,4,5$, the highest LOOCV $Q^2$ was achieved at $k=3$ ($R^2=0.733$, $Q^2=0.631$, RMSE $=0.279$, $n=21$), which does not meet the pass threshold.

The best $k=3$ model is:
\begin{equation}
\ln(k_{\mathrm{obs}})_{\mathrm{MeCN}} = 25.86\,x_1 + 0.00\,x_2 + 38.81\,x_3 - 4.75,
\end{equation}
where $x_1$ denotes the HOMO energy (\texttt{homo\_hartree}),
$x_2$ denotes the SCF total energy (\texttt{scf\_energy\_hartree}), i.e., the final converged electronic energy reported by the SCF procedure (an output descriptor rather than a user-specified input parameter),
and $x_3$ denotes the atomic charge at site $b$ (\texttt{q\_b}).

\begin{center}
\casefig{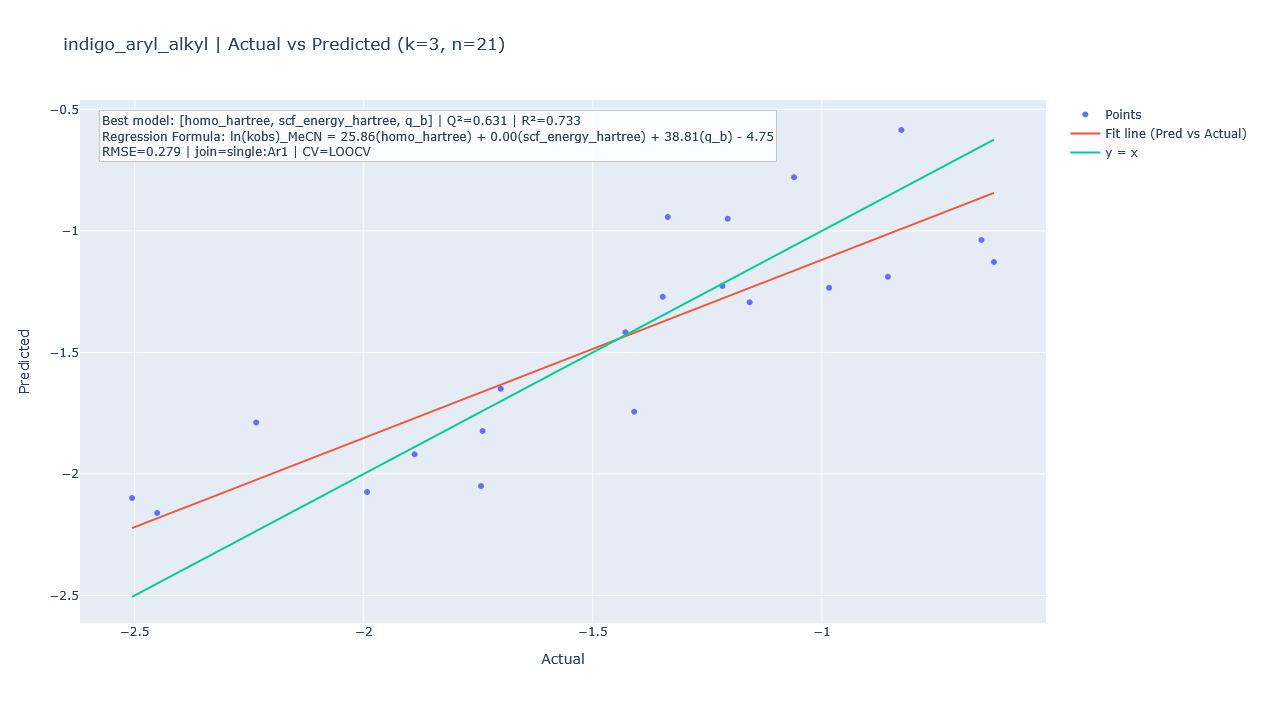}

\vspace{0.1em}
\casefig{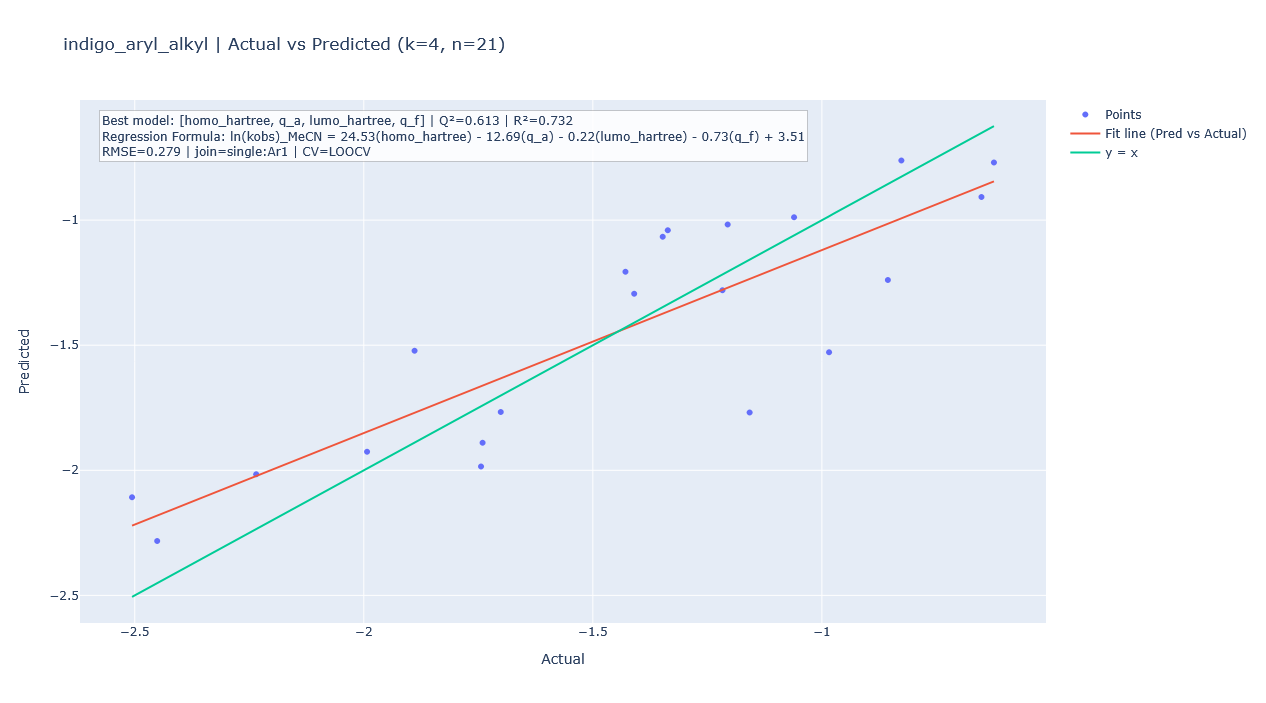}

\vspace{0.1em}
\casefig{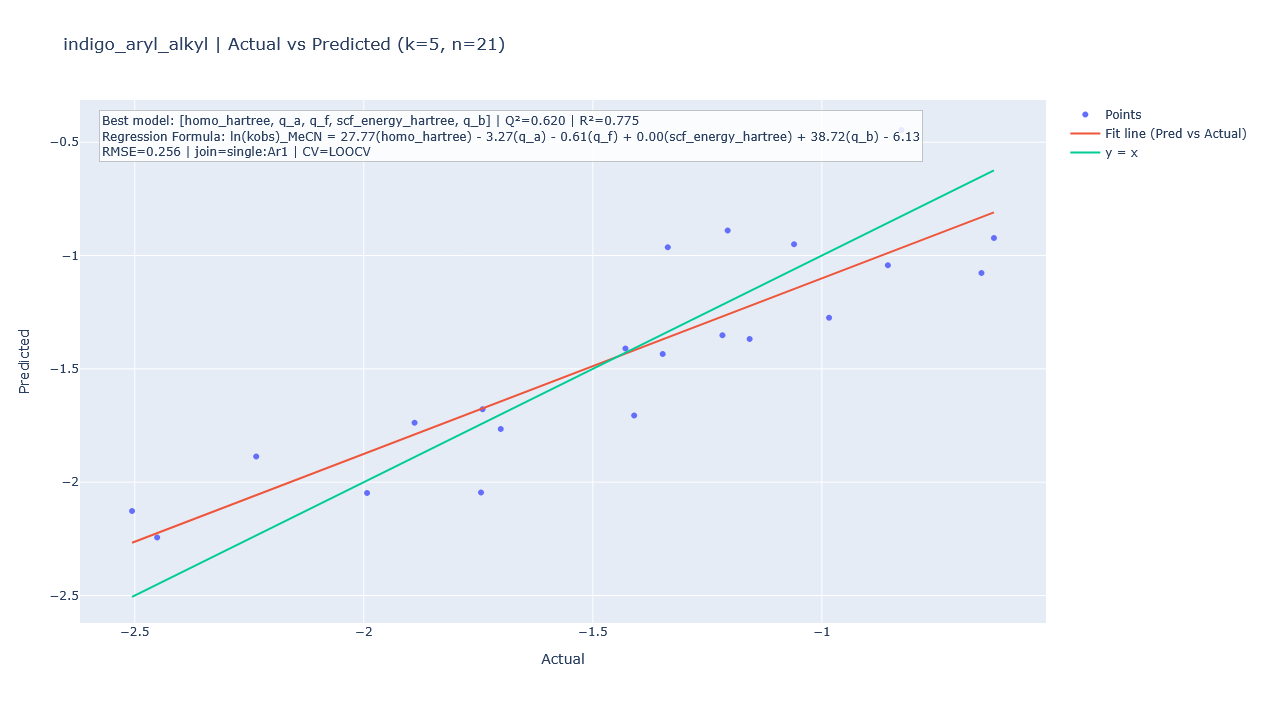}

\captionof{figure}{Indigo Aryl--Alkyl regression plots under \texttt{single:Ar1} join.
(a) $k=3$: $n=21$, $R^2=0.733$, $Q^2=0.631$, RMSE $=0.279$.
(b) $k=4$: $n=21$, $R^2=0.732$, $Q^2=0.613$, RMSE $=0.279$.
(c) $k=5$: $n=21$, $R^2=0.775$, $Q^2=0.620$, RMSE $=0.256$.}
\label{fig:indigo_mono}
\end{center}
\subsection{Thermal Back-isomerization of \emph{N},\emph{N}$'$-diarylindigo Photoswitches}

Finally, we considered bis-aryl indigo derivatives featuring two independently variable aryl substituents (Figure~\ref{fig:scheme4}).
This Indigo Diaryl dataset uses a \textbf{pair-join} scheme (\texttt{pair:Ar1+Ar2}) with target $\ln(k_{\mathrm{obs}})$.
In contrast to the previous three case studies, this dataset produces models that satisfy the pass criteria. At $k=3$, we already obtain a passing model with strong generalization ($R^2=0.967$, $Q^2=0.950$, RMSE $=0.338$, $n=13$).
Higher-$k$ variants further improve numerical metrics but may introduce less interpretable terms (e.g., an \texttt{Entry} index); therefore we use the $k=3$ model as the primary reported equation.

The primary ($k=3$) model is:
\begin{equation}
\ln(k_{\mathrm{obs}}) = 49.79\,x_1 - 0.25\,x_2 - 0.02\,x_3 + 477.87,
\end{equation}
where $x_1$ denotes the Ar$_2$ LUMO energy (\texttt{Ar2\_LUMO}),
$x_2$ denotes $\nu(\mathrm{C{=}O})$ of Ar$_1$ (\texttt{Ar1\_v\_C=O}),
and $x_3$ denotes $I(\mathrm{C{=}O})$ of Ar$_1$ (\texttt{Ar1\_I\_C=O}).

\begin{center}
\casefig{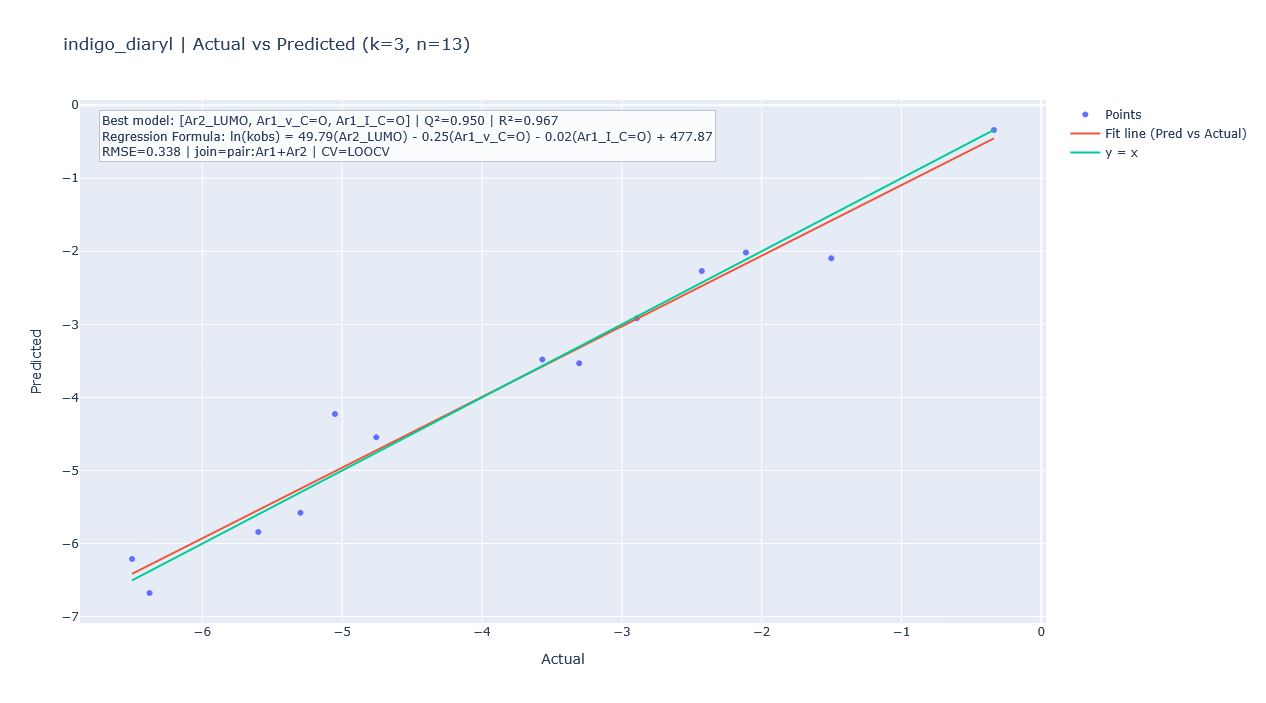}

\vspace{0.1em}
\casefig{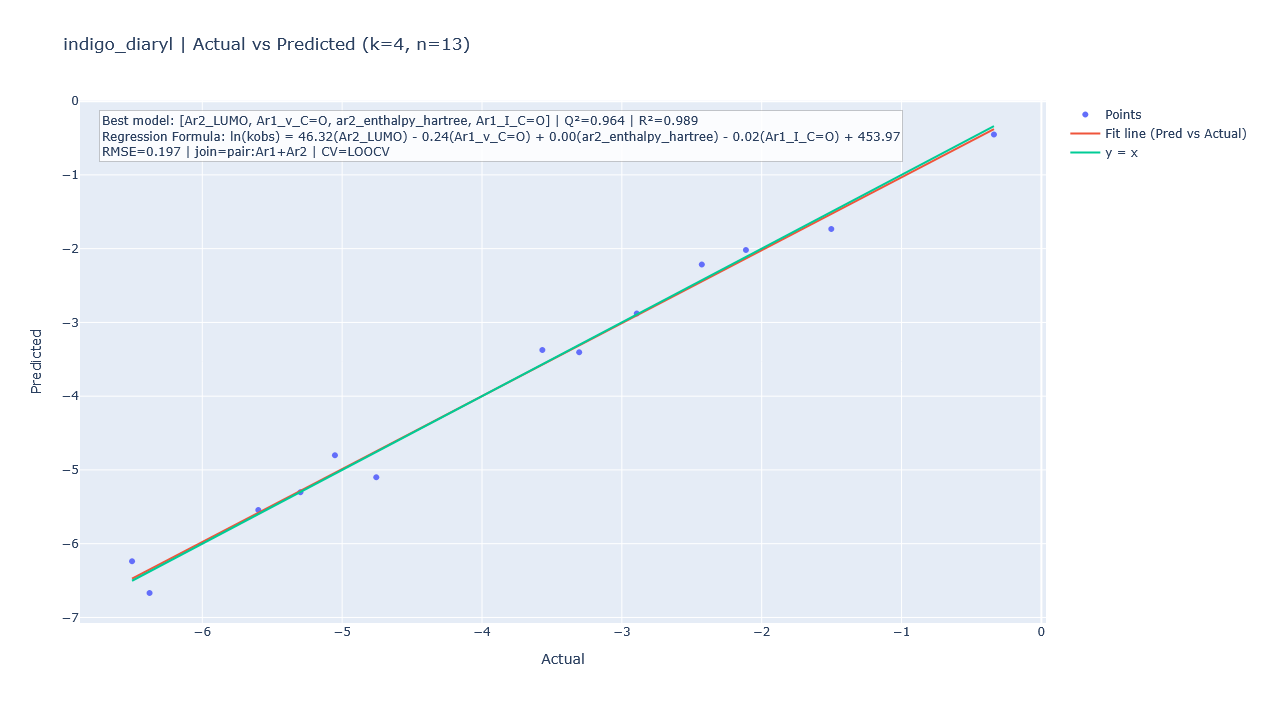}

\vspace{0.1em}
\casefig{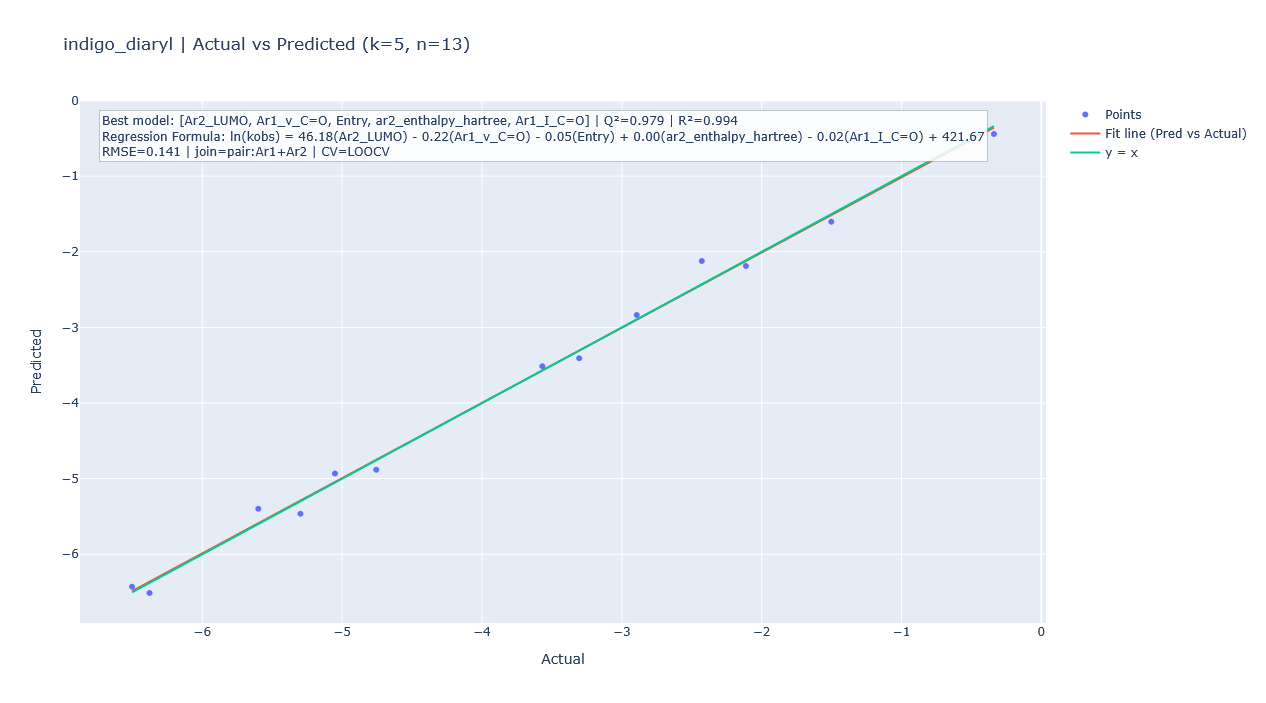}

\captionof{figure}{Indigo Diaryl regression plots under \texttt{pair:Ar1+Ar2} join.
(a) $k=3$ (PASS): $n=13$, $R^2=0.967$, $Q^2=0.950$, RMSE $=0.338$.
(b) $k=4$ (PASS): $n=13$, $R^2=0.989$, $Q^2=0.964$, RMSE $=0.197$.
(c) $k=5$ (PASS): $n=13$, $R^2=0.994$, $Q^2=0.979$, RMSE $=0.141$.}
\label{fig:indigo_bis}
\end{center}

\begin{landscape}
\begin{table}[t]
\caption{Summary of four case studies and their best-performing subset-selected regression models (selected by LOOCV $Q^2$), including join mode, target, dataset size, key descriptors, and performance metrics ($R^2$, $Q^2$, RMSE).}
\label{tab:summary}
\centering
\footnotesize
\setlength{\tabcolsep}{5pt}
\renewcommand{\arraystretch}{1.15}

\begin{tabular}{@{} l l l c c c
  >{\RaggedRight\arraybackslash}p{0.34\textwidth}
  >{\RaggedRight\arraybackslash}p{0.23\textwidth} @{}}
\toprule
Case Study & Join Mode & Target & $n_{\mathrm{raw}}$ & $n_{\mathrm{eff}}$ & $k_{\mathrm{best}}$ & Key Descriptors (Best) & Regression Performance (Best) \\
\midrule
Azoarene & \code{pair:Ar1+Ar2} & $\ln(k_{\mathrm{obs}})$ & 26 & 26 & 4 &
\makecell[l]{\code{Ar2\_homo\_hartree}\\ \code{Ar2\_scf\_energy\_hartree}\\ \code{Ar1\_q\_a}\\ \code{Ar2\_q\_g}} &
\makecell[l]{$R^2=0.647$\\ $Q^2=0.495$\\ RMSE $=2.63$} \\
\midrule
Heck-Boronic-Acids & \code{single:Ar} & $\Delta\Delta G$ & 17 & 17 & 5 &
\makecell[l]{\code{q\_e}\\ \code{q\_a}\\ \code{q\_d}\\ \code{q\_b}\\ \code{q\_f}} &
\makecell[l]{$R^2=0.889$\\ $Q^2=0.745$\\ RMSE $=0.223$} \\
\midrule
\emph{N}-aryl-\emph{N}$'$-alkylindigo & \code{single:Ar1} & $\ln(k_{\mathrm{obs}})_{\mathrm{MeCN}}$ & 21 & 21 & 3 &
\makecell[l]{\code{homo\_hartree}\\ \code{scf\_energy\_hartree}\\ \code{q\_b}} &
\makecell[l]{$R^2=0.733$\\ $Q^2=0.631$\\ RMSE $=0.279$} \\
\midrule
\emph{N},\emph{N}$'$-diarylindigo & \code{pair:Ar1+Ar2} & $\ln(k_{\mathrm{obs}})$ & 13 & 13 & 3 &
\makecell[l]{\code{Ar2\_LUMO}\\ \code{Ar1\_v\_C=O}\\ \code{Ar1\_I\_C=O}} &
\makecell[l]{$R^2=0.967$\\ $Q^2=0.950$\\ RMSE $=0.338$} \\
\bottomrule
\end{tabular}

\end{table}
\end{landscape}

\section{Conclusion}
The \textsc{DFTDescriptorPipeline} provides a convenient workflow for establishing structure--property and structure--reactivity relationships through computational chemistry and linear free-energy analysis. By integrating fully automated descriptor extraction, flexible feature engineering, and multivariate linear regression modeling into a unified open-source platform, this work greatly accelerates the translation of quantum-chemical calculations into actionable, data-driven insights.

A key strength of this pipeline is its extensibility. Every module is independently testable, and new descriptor types can be flexibly accommodated at the discretion of users without substantial code modification. This modularity empowers researchers to quickly adapt the workflow to novel molecular classes, reaction types, or emerging physical descriptors, ensuring the tool remains relevant as chemical knowledge and computational methods advance.

In addition, by supporting high-throughput, fully reproducible modeling with minimal user intervention, this approach significantly lowers the barrier for non-specialists to apply quantum-chemical data in experimental and industrial settings. Its design also facilitates integration with laboratory automation, database mining, and machine-learning pipelines, making it a foundation for future closed-loop discovery platforms in catalysis, materials science, and medicinal chemistry.

The methodology could be further extended to support more sophisticated statistical and machine-learning models, automated uncertainty quantification, and real-time feedback between quantum calculations and experiment. Furthermore, interfacing this workflow with platforms for molecular structure input and quantum-chemical computation is expected to increase scalability and usability for non-specialists. We anticipate its broad adoption and continual growth as a community-driven platform for data-centric chemistry.

\begin{acknowledgement}
D. C.-Y. H. acknowledges support from WPI-ICReDD (Hokkaido University), Oklahoma State University, and JSPS (JP23 K13734). Y.-C. H. and D. Y.-U. T. acknowledges financial support from National Taiwan Normal University (NTNU).
\end{acknowledgement}

\section*{Associated Content}
Code examples and the job/log files for the four case studies (azoarenes, redox-relay Heck coupling, \emph{N}-aryl-\emph{N}$'$-alkylindigos, and \emph{N},\emph{N}$'$-diarylindigos) are openly available in the project repository (\url{https://github.com/peculab/DFTDescriptorPipeline}).

\bibliography{achemso-demo}

\end{document}